\documentclass[showpacs,preprintnumbers,amsmath,amssymb,12pt]{revtex4}
\usepackage{bm}
\begin{document}

\title[]{Tidal fields on brane worlds}

\author{Edgard Casal de Rey Neto}

\email{{ecrneto@ig.com.br}}
\affiliation{Instituto Tecnol\'ogico de Aeron\'autica - Divis\~ao de
F\'{\i}sica Fundamental \\ Pra\c ca Marechal Eduardo Gomes 50, S\~ao Jos\'e dos
Campos, 12228-900 SP, Brazil}

\date{06/11/2004}

\begin{abstract}

We write out the geodesic deviations that take place in a $d\geq 4$ dimensional brane world subspace  of  a higher dimensional spacetime by splitting out the brane and the extra space dynamical quantities from a global metric spacetime of dimension $D\geq 5$. The higher dimensional dynamical quantities are projected onto two orthogonal subspaces, where one of which is identified with a $(d-1)$-brane. This is done by using some technics of the conventional submanifold theory of the Riemannian geometry, applied to pseudo-Riemannian spaces. Using the splitting technic, we obtain the tidal field on $(d-1)$ branes with an arbitrary number of non compact extra dimensions. Later, we analise the geodesic deviations seen by an ordinary observer in a $d=4$ dimensional spacetime and show that deviations from general relativity tidal field due to the existence of the extra dimensions can appear because, (i) - the dependence of the indunced metric on the brane with the extra coordinates and (ii) - deviations of the higher dimensional spacetime metric from spherical symmetry. 
\end{abstract}

\maketitle

\newpage

\section{Introduction}

It is an old idea that spacetime may have more than four dimensions. This idea was first introduced in spacetime physics by Kaluza  at 1920s, by constructing a field theory in a five dimensional spacetime of the gravitational and electromagnetic interactions. To explain the non observation of the extra dimensions, the Kaluza original idea was to impose the so called cylindric conditions, in which the four dimensional metric and the all physical four dimensional quantities are independent of the extra coordinate. A distinct mechanism to lead to an effective four dimensional gravity, was proposed by Klein also in the twenty years, by enable the extra dimensional coordinates range in  small compact sub manifolds with characteristic sizes of order of Planck scale, leading to an effective four dimensional gravity in observable distances, or scales. For a long time, the small compact extra dimensions was the paradigm of higher-dimensional physics, being the basis of the development of the most of string theories and, the only generally acceptable way to get a four dimensional physics form a higher dimensional spacetime~\cite{OW1}. A different idea is that our universe may be a thin membrane in a higher dimensional bulk spacetime. In such picture of the universe, is called  brane world, or brane-universe, the extra dimensions may be non compact or even infinite.  

The early works on brane models have appeared in the eighteens~\cite{earlyw}. However, only after the works~\cite{ADD,RS1}, brane world models have become popular. In such models, the gravity can propagate in all dimensions while the matter is confined on a spacetime subspace of dimension lower then the dimension of the global spacetime, or bulk. The large or infinite extra dimensions play an important role in solve problems as smallness of cosmological constant, the origin of the hierarchy between gravity and standard interactions, etc. A natural mechanism for matter localization on the brane is needed in such models. Here, we do not study the localization mechanisms, we only assume that general matter are localized on branes and, in particular, the ordinary, or standard model matter, is localized on a four dimensional membrane that is identified with the world where the  observations can taken place. The issue of matter localization is discussed in several works in the literature as, for instance, in~\cite{GS1,GM1,DGS1} and references cited therein.

In the preset paper we are interested on the possibility of use gravitational wave antennas to search for observable effects of extra spacetime dimensions in such models. The gravitational wave antennas are building to be sensitive to local tidal fields,  produced by weak dynamical deviations of the local spacetime metric from the spacetime background, which is generally assumed to be flat. Therefore, to study gravitational wave antenna sensitivity we must begin by write the geodesic deviation equations in a suitable form to be applicable to a four dimensional observer. 

We obtain the geodesic deviations in brane world subspaces of a higher dimensional spacetime, by projecting the higher dimensional dynamical quantities in two orthogonal subspaces, one of which is identified with a $d$ dimensional membrane. We write the tidal field that is seen by an observer in four dimensions is given by the geodesic deviations on a four dimensional subspace, a 3-brane. In a real antenna that can be constructed in earth or space laboratories, the test particles are confined to move only on a 3-brane subspace, this assumption reduce the right hand side terms that appear in the general geodesic deviation equation for a classical antenna. Even with such reduction, the existence of the extra dimensions, or of the co-dimension space, can modify the the gravitational tidal field relatively to the classical general relativity formulae. 

We assume a $D\geq 5$ dimensional metric spacetime in which the weak equivalence principle,  as stated by~\cite{Wein},  holds. Thus, we have the usual relation between the Christoffel symbols and the metric of the global $D$ dimensional spacetime, or bulk. 

In the section~\ref{sec2}, we present the formalism which enable us to make a decomposition of a global $D$ dimensional spacetime in two orthogonal subspaces and define the space of the projected higher dimensional coordinates. In the section~\ref{sec3}, we obtain the geodesic deviations on the $(d-1)$-brane world subspaces and concentrate on the case of a free test particles localized on the brane hypersurface. In the section~\ref{sec4}, we give some immediate conclusions that can be traced from the effective tidal field by considering some examples of general bulk metrics.

\section{The splitting of the global spacetime in to two orthogonal subspaces}
\label{sec2}

Let ${\cal M}$ be a $D$ dimensional pseudo-Riemannian manifold with metric tensor $G_{AB}$ and local coordinates $X^A (A,B=0,...,D-1)$. We call ${\cal M}$ the global spacetime. Let ${\cal X}$ be a $d$ dimensional sub-manifold of ${\cal M}$ with signature $(-1,1,1,...)$ and local coordinates $x^\mu$ $(\mu=0,...,d-1)$. Then, parametric equation of ${\cal X}$ is
\begin{equation}
\label{xsup}
X^A=X^A(x^0,...,x^{d-1}), \quad A=0,...,D-1.
\end{equation}

Now, let ${\cal Y}$ be a $k=D-d$ dimensional sub-manifold of $M$  with local coordinates $y^a (a=d,...,D)$ and arbitrary signature. The parametric equation of ${\cal Y}$ is
 
\begin{equation}
\label{xsup}
X^A=X^A(y^d,...,y^{D-1}), \quad A=0,...,D-1.
\end{equation}
The tangent vector in the coordinate directions on ${\cal X}$ an d${\cal Y}$, namely, the coordinate basis on ${\cal X}$ and ${\cal Y}$ subspaces are given, respectively, by
\begin{equation}
\label{tvbX}
\frac{\partial}{\partial x^\mu}=e^A_{(\mu)}\frac{\partial}{\partial X^A}, \quad e^A_{(\mu)}=\frac{\partial X^A}{\partial x^\mu},
\end{equation}
and 
\begin{equation}
\label{tvbY}
\frac{\partial}{\partial y^a}=e^A_{(a)}\frac{\partial}{\partial X^A}, \quad e^A_{(a)}=\frac{\partial X^A}{\partial y^a},
\end{equation}
is the expansion of the tangent vector basis of the ${\cal Y}$ sub-space.
One can also expand the coordinate basis of the global spacetime, $\{\partial/\partial X^A\}$, in terms of the basis $\{\partial/\partial x^\mu\}$ and $\{\partial/\partial y^a\}$, namely
\begin{equation}
\label{tvMxy}
\frac{\partial}{\partial X^A}=e^{(\mu)}_A\frac{\partial}{\partial x^\mu}+e^{(a)}_A\frac{\partial}{\partial y^a}, \quad
{\rm where}\quad
e^{(\mu)}_A=\frac{\partial x^\mu}{\partial X^A},\quad e^{(a)}_A=\frac{\partial y^a}{\partial X^A}.
\end{equation}
From~(\ref{tvMxy}), and the definitions~(\ref{tvbX}) and~(\ref{tvbY}) we get 
\begin{equation}
\label{kdel1}
e^{(\mu)}_Ae^B_{(\mu)}+e^{(a)}_Ae^B_{(a)}=\delta^B_A.
\end{equation}
One can also see that
\begin{equation}
\label{kdel2}
e^A_{(\mu)} e^{(\nu)}_A=\delta^\nu_\mu,\quad e^A_{(a)}e^{(b)}_A=\delta^b_a.
\end{equation}
The induced metric tensors on ${\cal X}$ and ${\cal Y}$ are given respectively by
\begin{equation}
\label{metricX}
g_{\mu\nu}=e^A_{(\mu)} e^B_{(\nu)} G_{AB},\quad\mu,\nu=0,...,d-1,
\end{equation}
and
\begin{equation}
\label{metricY}
g_{ab}=e^A_{(a)} e^B_{(b)}G_{AB},\quad a,b=d,...,D-1.
\end{equation}
The inner product of pairs of vectors in each subspace are defined in the usual way from their respective induced metrics.
Naturally,
\begin{equation}
g_{\mu a}=e^A_{(\mu)}e^B_{(a)}G_{AB}
\end{equation}
provides the inner product of a vector $x^\mu\in{\cal X}$ with a vector $y^a\in{\cal Y}$.
When $g_{\mu a}=0$, the subspaces ${\cal X}$ and ${\cal Y}$ are orthogonal one each orther.
The ${\cal X}$ sub-space is called a $(d-1)$-brane and the ${\cal Y}$ sub-space is called the extra, or codimension, space. The $(d-1)$ brane is a $d$ dimensional spacetime with one time-like and $d-1$ space-like dimensions.

All the $D$ dimensional quantities can be expanded in terms of the quantities defined in the $X$ and $Y$ sub-spaces. In particular,
 dimensional contravariant vector $X^A$ can be expanded in terms of the contravariant vectors in the mutually orthogonal sub-spaces as
\begin{equation}
X^A=e^A_{(\mu)} x^\mu+e^A_{(a)} y^a.
\end{equation}
Also,  any $D$ dimensional covariant vector $X_A$ can be expanded as
\begin{equation}
X_A=e^{(\mu)}_A x_\mu+e^{(a)}_A y_a,
\end{equation}
where
\begin{equation}
\label{subcoord}
x^\mu=e^{(\mu)}_A X^A\;\;,\;\; y^a=e^{(a)}_A X^A\;\;\;;\;\;\;x_\mu=e_{(\mu)}^A X_A\;\;,\;\; y_a=e_{(a)}^A X_A.
\end{equation}
The metric tensor of the global spacetime can be expanded as
\begin{equation}
\label{gmetricexp}
G_{AB}=e^{(\mu)}_Ae^{(\nu)}_Bg_{\mu\nu}+\left(e^{(\mu)}_Ae^{(a)}_B+e^{(a)}_Ae^{(\mu)}_B\right)g_{\mu a}+e^{(a)}_Ae^{(b)}_Bg_{ab}.
\end{equation}

Let us supose that for a given choice of coordinates $x^\mu$ and $y^a$ we have $g_{\mu a}=0$. Then, in such coordinates the global spacetime line element reduces to the form a higher dimensional generalization of the Randall-Sundrum line element~\cite{RS1} with $g_{\mu\nu}$ plaing the role of induced metric on the brane and and $g_{ab}$ the induced metric on the codimension space. The subspaces ${\cal X}$ and ${\cal Y}$ are said to be orthogonal one each other. Such decomposition of the global spacetime in two orthogonal submanifolds can be realized whereas one can find projectors $e^A_{(\mu)}$ and $e^A_{(a)}$ shch that
\begin{equation}
\label{ortcond}
g_{\mu a}=G_{AB}e^A_{(\mu)}e^B_{(a)}=0.
\end{equation}
The relation~({\ref{ortcond}) is called orthogonality relation.
When the orthogonality condition holds, the algebric relations are greately simplified. The induced metrics $g_{\mu\nu}$ and $g_{ab}$ can be used to raise and lower the indices of each subspace individually, namely
\begin{equation}
\label{rlindices}
e^{(\mu)}_A=g^{\mu\nu}G_{AB}e^B_{(\nu)},\quad e^{(a)}_A=g^{ab}G_{AB}e^B_{(b)}.
\end{equation}
Also,
\begin{equation}
\label{gmetricexpinv}
G^{AB}=e^A_{(\mu)} e^B_{(\nu)} g^{\mu\nu}+e^A_{(a)}e^B_{(b)}g^{ab},
\end{equation}
\begin{equation}
\label{indmetricinv}
g^{\mu\nu}=e^{(\mu)}_Ae^{(\nu)}_BG^{AB},\quad g^{ab}=e^{(a)}_A e^{(b)}_BG^{AB}.
\end{equation}
and
\begin{equation}
\label{ortinv}
G^{AB}e^{(\mu)}_A e^{(a)}_B=0,
\end{equation}
where $G^{AB}$, $g^{\mu\nu}$ and $g^{ab}$ are the inverses of $G_{AB}$, $g_{\mu\nu}$ and $g_{ab}$, respectively.

We assume that the higher dimensional spacetimes that we shall study in this paper can be decomposible in two orthogonal submanifolds. That is, given a global higher dimensional spacetime manifold ${\cal M}$ one can find projectors which sitisfy~(\ref{ortcond}). The coordinates of the orthogonal submanifolds are just given by~(\ref{subcoord}).

Now, we improve our notation to a more compact one which will be useful in the computations of the next sections. We define the enuples of the $D$-dimensional vectors
\begin{equation}
\label{compnotd1}
e^{(B)}_A=(e^{(\mu)}_A,e^{(a)}_A),\quad e^A_{(B)}=(e^A_{(\mu)},e^A_{(a)}),
\end{equation}
where $\mu=0,...,d-1$, $a=d,...,D-1$, $A=0,...,D-1$.
According to the new definitions, the relations~(\ref{kdel1}) and~(\ref{kdel2}) are rewritten as
\begin{equation}
\label{kdel3}
e^C_{(A)}e^{(B)}_C=\delta^B_A,\quad e^{(A)}_Be^C_{(A)}=\delta^C_B.
\end{equation}
We also define a local $D$ dimensional metric $g_{(A)(B)}$ by
\begin{equation}
\label{localGAB1}
g_{(A)(B)}=e^{C}_{(A)}e^{D}_{(B)}G_{CD},
\end{equation}
or the converse relation:
\begin{equation}
\label{localGAB2}
G_{AB}=e^{(C)}_{A}e^{(D)}_{B}g_{(C)(D)},
\end{equation}
From the orthogonality relation~(\ref{ortcond}), one can see that $g_{(\mu)(a)}= 0$, even when $G_{\mu a}\neq 0$.

The quantities defined in~(\ref{compnotd1}), projects a $D$ dimensional vector of ${\cal M}$ on the subspaces ${\cal X}$ and ${\cal Y}$. Then, the $D$-dimensional spacetime ${\cal N}$ with coordinates $X^{(A)}=e^{(A)}_B X^B ((A)=0,...,D-1)$, is the space of the projections on ${\cal X}$ and ${\cal Y}$. Note that, $X^{(\mu)}=x^\mu$ and $X^{(a)}=y^a$.

We now write the relationship among the Christoffel symbols of ${\cal M}$ and ${\cal N}$. From~(\ref{localGAB2}), $\Gamma^A_{BC}$ can be expressed as
\begin{equation}
\label{GammaABC}
\Gamma^A_{BC}=e^A_{(A)}e^{(B)}_Be^{(C)}_C\Gamma^{(A)}_{(B)(C)}+e^A_{(A)}e^{(A)}_{B,C},
\end{equation}
where $e^{(A)}_{B,C}=\partial e^{(A)}_{B}/\partial X^C$.
Using~(\ref{kdel3}), one can show that
\begin{equation}
\label{GammaABCl}
\Gamma^{(A)}_{(B)(C)}=e^{(A)}_Ae^B_{(B)}e^C_{(C)}\Gamma^A_{BC}+e^{(A)}_Ae^A_{(B)|(C)},
\end{equation}
where the $\mid$ means the projected derivative defined by
\begin{equation}
\label{projder}
(...)_{\mid(B)}=(...)_{,\;B}e^B_{(B)}.
\end{equation}

\section{Geodesic deviations}
\label{sec3}

Let us now obtain an expression for geodesic deviations in brane world spacetimes. First of all we must write the geodesic equations in the ${\cal N}$ space. These equations was derived Ponce de Leon in~\cite{PL1} for the five dimensional Kaluza-Klein gravity. 

The geodesic equations on the spacetime ${\cal M}$, have the same form of the general relativity geodesic equation with the indexes of the four dimensional spacetime replaced by the $D$ dimensional indices, namley,
\begin{equation}
\label{geM}
\frac{d^2X^{A}}{dS^2}+\Gamma^{A}_{BC}\frac{dX^{B}}{dS}\frac{dX^{C}}{dS}=0.
\end{equation}
The geodesic equations in ${\cal N}$ are 

\begin{equation}
\label{geN}
\frac{d^2X^{(A)}}{dS^2}+\Gamma^{(A)}_{(B)(C)}\frac{dX^{(B)}}{dS}\frac{dX^{(C)}}{dS}=F^{(A)},
\end{equation}
where,
\begin{equation}
\label{Fdef}
F^{(A)}=G^{(A)(B)}U_{(a)}e^{(a)}_{A}(e^A_{(B)\mid(C)}-e^A_{(C)\mid(B)})U^{(C)},\quad a=d,...,D-1
\end{equation}
and $U^{(B)}=dX^{(B)}/dS$.
To arrive at~(\ref{geN}) one must note that
\begin{equation}
\label{dXs}
\frac{dX^A}{dS}=\frac{\partial X^A}{\partial X^{(B)}}\frac{dX^{(B)}}{dS}=e^A_{(B)}U^{(B)}.
\end{equation}
Therefore,
\begin{equation}
\label{accXA}
\frac{d^2X^A}{dS^2}=e^A_{(B)\mid(C)}+e^A_{(B)}U^{(B)}.
\end{equation}
Then, insert this and~(\ref{localGAB2}) in the geodesic deviation equation of ${\cal M}$ space~\cite{PL1}. 

The right hand side of equation~(\ref{geN}) have contributions of the so called extra force and of the $G_{\mu a}$ components of the bulk metric. By choosing the free index of~(\ref{geN}) to run only on the ${\cal X}$ space we obtain the free accelerations of a test particle, as it can be seen form a local frame in ${\cal X}$.
A fundamental point to be noted is that $F^{(A)}=0$, for test particles localized on the brane ($U_{(a)}=0$) and the geodesic motion on ${\cal X}$ is given by the classical geodesic equation on this sub-space~\cite{Y1}.

Before the derivation of the geodesic deviations on ${\cal X}$, we must taken into account some important features, that.
Firstly, let us consider the bulk spacetme line element
\begin{equation}
\label{dSglob}
dS^2=G_{AB}dX^AdX^B.
\end{equation}
Defining $dx^\mu=e^{(\mu)}_A dX^A$ and $dy^a=e^{(a)}_A dX^A$ we can write
\begin{equation}
\label{dSsep}
dS^2=g_{\mu\nu}dx^\mu dx^\nu+g_{ab}dy^ady^b.
\end{equation}
If we want to obtain the geodesic motion of a test particle as it is seen by an observer on ${\cal X}$, we must write the geodesic equations in terms of a new affine parameter that can be linearly related to the proper time measured on the ${\cal X}$ sub-space. Then, by following~\cite{Y1} we define a new affine parameter $s$ such that
\begin{equation}
\label{epslon4def}
\epsilon_d=g_{\mu\nu}\frac{dx^\mu}{ds}\frac{dx^\nu}{ds},
\end{equation}
where $\epsilon_d=-1,0,1$ respectively for timelike, null and spacelike geodesics on ${\cal X}$ and $s$ is a differentiable function of $S$. Inserting $dS=\frac{dS}{ds}ds$ in~(\ref{geN}) we obtain the geodesic deviations of the ${\cal N}$ coordinates in terms of the affine parameter $s$.
\begin{equation}
\label{geNs}
\frac{du^{(A)}}{ds}+\Pi u^{(A)}+\Gamma^{(A)}_{(B)(C)}u^{(B)}u^{(C)}=f^{(A)},
\end{equation}
where $u^{(A)}=dX^{(A)}/ds$,
\begin{equation}
\label{ch1def1}
\Pi=\left(\frac{ds}{dS}\right)^{-1}\frac{d}{ds}\left(\frac{ds}{dS}\right),
\end{equation}
and
\begin{equation}
\label{fdef}
f^{(A)}=G^{(A)(B)}u_{(a)}e^{(a)}_{A}(e^A_{(B)\mid(C)}-e^A_{(C)\mid(B)})u^{(C)}.
\end{equation}

Now, we can follow the standard steps on the derivation of the geodesic deviations such as in~\cite{Wein}, to write out the geodesic deviations in the projected space ${\cal N}$. Given a small displacement $\delta X^{(A)}$ in ${\cal N}$ we have
\begin{eqnarray}
\label{gdN1}
\frac{D^2\delta X^{(A)}}{ds^2}&=&-{R^{(A)}}_{(B)(C)(D)}u^{(B)}\delta X^{(C)}u^{(B)}-(u^{(A)}\Pi_{\mid(B)}-f^{(A)}_{\mid(B)})\delta X^{(B)}\nonumber \\&-&\Pi\frac{d\delta X^{(A)}}{ds}
+\Gamma^{(A)}_{(B)(C)}\delta X^{(B)}f^{(C)}.
\end{eqnarray}
The world line deviations in each of the two orthogonal subspaces, a $(d-1)$-brane and the extra space, can be obtained from~(\ref{gdN1}), by choosing the free index to run in one of these subspaces.

The equation that gives the geodesic deviations in the subspace ${\cal X}$, is obtained by  setting the free index $A=\mu$ in~(\ref{gdN1}), namely
\begin{eqnarray}
\label{gdmu}
\frac{D^2\delta X^{(\mu)}}{ds^2}&=&-{R^{(\mu)}}_{(\nu)(\rho)(\sigma)}u^{(\nu)}\delta X^{(\rho)} u^{(\sigma)}-\Pi\frac{d\delta X^{(\mu)}}{ds}-(u^{(\mu)}\Pi_{\mid(\nu)}-{f^{(\mu)}}_{\mid(\nu)})\delta X^{(\nu)}\nonumber \\
&-&(u^{(\mu)}\Pi_{\mid (a)}-{f^{(\mu)}}_{\mid (a)})\delta X^{(a)}
+\Gamma^{(\mu)}_{(\rho)(\sigma)}\delta X^{(\rho)} f^{(\sigma)}+\Gamma^{(\mu)}_{(\rho) (a)}\delta X^{(\rho)} f^{(a)}\nonumber \\
&+&\Gamma^{(\mu)}_{(a) (b)}\delta X^{(a)} f^{(b)}-{R^{(\mu)}}_{(a)(\rho)(\sigma)}u^{(a)}\delta X^{(\rho)} u^{(\sigma)}-{R^{(\mu)}}_{(\nu) (a)(\sigma)}u^{(\nu)}\delta X^{(a)} u^{(\sigma)}\nonumber \\&-&{R^{(\mu)}}_{(a)(b)(\sigma)}u^{(a)}\delta X^{(b)} u^{(\sigma)}-{R^{(\mu)}}_{(a)(b)(c)}u^{(a)}\delta X^{(b)} u^{(c)},
\end{eqnarray}
where we have defined  $\delta X^{(\mu)}=e^{(\mu)}_A\delta X^A$ and $\delta X^{(a)}=e^{(a)}_A\delta X^A$.

The scalar $\Pi$ have different dependencies on the velocity of the test particles for null and non null geodesics in the global spacetime ${\cal M}$.
For non null bulk geodesic,
\begin{equation}
\label{leGAB}
\epsilon_{D}=G_{AB}\frac{dX^A}{dS}\frac{dX ^B}{dS},
\end{equation}
where $\epsilon_D=-1,1$. Without loss of generality, we can study only $\epsilon_D=-1$ case - timelike geodesics in ${\cal M}$. Then, using $dS=ds(dS/ds)$ in~(\ref{leGAB}) and, writing $G_{AB}$ in terms of the induced metrics $g_{\mu\nu}$ and $g_{ab}$ we have

\begin{equation}
\label{Pi1}
\Pi=\frac{u_{(a)}du^{(a)}/ds}{(-\epsilon_d+u_{(a)}u^{(a)})^2}\;,
\end{equation}
where $u_{(a)}=dy_{(a)}/ds$.
For null  bulk geodesics, $\epsilon_D=0$, splitting out the ${\cal X}$ and ${\cal Y}$ components of~(\ref{geNs}) and using~(\ref{leGAB}) one can show that
\begin{equation}
\label{Pi2}
\Pi=-u_{(a)}{\Gamma^{(a)}}_{(b)(c)}u^{(b)}u^{(c)}-g_{ab\mid(\mu)}u^{(a)} u^{(b)}u^{(\mu)}+
\frac{1}{2}g_{\alpha\beta\mid(a)}u^{(\alpha)} u^{(\beta)} u^{(a)}.
\end{equation}
The above results are derived in~\cite{PL1,Y1} for braneworlds with one extra-dimension.

If the test particle is localized on the ${\cal X}$ subspace, then $u^{(a)}=0$, and the equation of geodesic deviations in ${\cal X}$ are given by~(\ref{gdmu}), with $u^{(a)}=0$. As can be seen from~(\ref{fdef}), (\ref{gdmu}), (\ref{Pi1}) and~(\ref{Pi2}), only the first term of the right hand side do not vanish when $u^{(a)}=0$. Thus, the geodesic deviations becomes
\begin{equation}
\label{gdmu2}
\frac{D^2\delta X^{(\mu)}}{ds^2}=-{R^{(\mu)}}_{(\nu)(\rho)(\sigma)}u^{(\nu)}\delta X^{(\rho)} u^{(\sigma)},
\end{equation}
where ${R^{(\mu)}}_{(\nu)(\rho)(\sigma)}$ is given by~\cite{YJY1},

\begin{equation}
\label{GC1}
{R^{(\mu)}}_{(\nu)(\rho)(\sigma)}={R^A}_{BCD}e^{(\mu)}_Ae^B_{(\nu)} e^C_{(\rho)} e^D_{(\sigma)}+\nabla_{(\rho)} e^{(\mu)}_A\nabla_{(\sigma)} e^A_{(\nu)}-\nabla_{(\sigma)} e^{(\mu)}_A\nabla_{(\rho)} e^A_{(\nu)},
\end{equation}
where the covariant derivatives of $e^{(\mu)}_A$ and $e^A_{(\mu)}$ are defined by
\begin{equation}
\label{dcovmuA}
\nabla_{(\rho)} e^{(\mu)}_A=e^{(\mu)}_{A\mid(\rho)}-\Gamma^B_{CA}e^{(\mu)}_B e^C_{(\rho)}+e^{(\sigma)}_A\Gamma^{(\mu)}_{(\rho)(\sigma)}
\end{equation}
and
\begin{equation}
\label{dcovAmu}
\nabla_{(\rho)} e^A_{(\mu)}=e^A_{(\mu)\mid(\rho)}-\Gamma^{(\sigma)}_{(\rho)(\mu)}e^A_{(\sigma)} +e^B_{(\rho)} e^C_{(\mu)}\Gamma^A_{BC}.
\end{equation}
The ${\cal X}$ subspace Christoffel symbols in~(\ref{dcovmuA}) and~(\ref{dcovAmu}), can be expressed in terms of the higher dimensional Christoffel symbols $\Gamma^A_{BC}$ by using~(\ref{GammaABCl}).

\section{conclusions}
\label{sec4}
For $d=4$, the equation~(\ref{GC1}) tell us how the tidal field measured in the ordinary spacetime dimensions can deviate from the general relativity prediction due to the existence of the extra dimensions. To interpret the terms of the right hand side of~(\ref{GC1}), we consider as a first case a global spacetime in which the metric tensor is such that $G_{\mu a}=0$, $\mu=0,...,d-1$ and $a=d,...D-d$. It means that the orginal spacetime ${\cal M}$ satisfy the orthogonality condition. Note that all higher dimensional sherically symmetric spacetimes fall in this  category~\cite{DDB1,SDL1}. When $G_{\mu a}=0$, the higher dimensional line element can be writeen as
\begin{equation}
\label{dSGort}
dS^2=G_{\mu\nu}dX^\mu dX^\nu+G_{ab}dX^adX^b.
\end{equation}
Therefore, one can choose the projectors $e^{(B)}_A=\delta^{(B)}_A$. With this choice, the least two terms of~(\ref{GC1}) vanishes and we have,
\begin{equation}
\label{Rcorr}
{R^{(\mu)}}_{(\nu)(\rho)(\sigma)}={{\cal R}^{(\mu)}}_{(\nu)(\rho)(\sigma)}-\frac{1}{2}g^{\mu\alpha} g_{\alpha[\rho,a}{g_{\sigma]\nu}}^{,a},
\end{equation}
where ${{\cal R}^{(\mu)}}_{(\nu)(\rho)(\sigma)}$ is the Riemann tensor constructed from $g_{\mu\nu}$ and  $2[\mu\nu]=\mu\nu-\nu\mu$. The least term of~(\ref{Rcorr}) tell us that, when the induced metric on ${\cal X}$, namely, $g_{\mu\nu}=G_{\mu\nu}$, depends on the extra coordinates $y^a$, there is a extra term in the geodesic deviations equation given by
\begin{equation}
\label{extterm1}
-\frac{1}{2}g^{\mu\alpha} g_{\alpha[\rho,a}{g_{\sigma]\nu}}^{,a}u^\nu\delta X^{(\rho)} u^{(\sigma)}.
\end{equation}
Therefore we can conclude that, if the metric $g_{\mu\nu}$ do not depend on extra coordinates, which correspond to the Kaluza cylindrical condition, the geodesic deviations are the same that those predicted in the four dimensional general relativity. In such case, the tidal field on the {\cal X} subspace cannot be affected in any way by the existence of the extra dimensions.

The full content of the equation~(\ref{GC1}) are active only when the original metric tensor of the global spacetime is such that $G_{\mu a}\neq 0$. In such cases, the projectors that appear in the right hand side of~(\ref{GC1}) cannot be made equal to the Kroneker delta for all  values of the indexes $A$ and $\mu$. Thus, the terms involving the derivatives of the projectors on the right hand side of~(\ref{GC1}) can give rise to deviations of the geodesic deviations from the general relativistic prediction. Examples of $D$ dimensional metric spacetimes for which $G_{\mu a}\neq 0$  are given by the higher dimensional gravitational waves in weak field approximation~\cite{CDL1p}. The quadrupole nature of the weak field gravitational waves produced by slow motion astrophysical sources implies that $G_{\mu a}\neq 0$. This feature will lead to nontrivial projectors $e^{(\mu)}_A$ in equation~(\ref{GC1}) that can bring the content of the higher dimensions to the lower dimensional tidal field on the brane. Since gravitational radiation produces small tidal field oscilations, the result obtained in the previous section, in particular the equations~(\ref{gdmu}) and~(\ref{GC1}), show that the gravitational wave antennas will be useful instruments to be used to obtain observational constraints on  higher-dimensional models.

\end{document}